\documentclass[prl,twocolumn]{revtex4}

\usepackage{graphicx}
\usepackage{epsfig}

\begin{document}

\title{Negative Differential Resistivity and Positive Temperature Coefficient of Resistivity effect in the
diffusion limited current of ferroelectric thin film capacitors}
\author{M. Dawber}
\affiliation{Dept of Earth Sciences, University of Cambridge,
Downing St, Cambridge CB2 3EQ, UK} \altaffiliation{Present address
of M. Dawber: DPMC, University of Geneva, 24 Quai E.-Ansermet 1211
Geneva 4, Switzerland } \email{matthew.dawber@physics.unige.ch}
\author{J.F. Scott}
\affiliation{Dept of Earth Sciences, University of Cambridge,
Downing St, Cambridge CB2 3EQ, UK}

\begin{abstract}
We present a model for the leakage current in ferroelectric thin-
film capacitors which explains two of the observed phenomena that
have escaped satisfactory explanation, i.e. the occurrence of
either a plateau or negative differential resistivity at low
voltages, and the observation of a Positive Temperature
Coefficient of Resistivity (PTCR) effect in certain samples in the
high-voltage regime. The leakage current is modelled by
considering a diffusion-limited current process, which in the
high-voltage regime recovers the diffusion-limited Schottky
relationship of Simmons already shown to be applicable in these
systems.

\end{abstract}

\maketitle

For a number of years the problem of understanding leakage
currents in ferroelectric capacitors (or high dielectric constant
capacitors using ferroelectric materials just above their phase
transtion, e.g. barium strontium titanate (BST)) has been of
practical interest, and a large number of papers have been
published on the subject, a review of which can be found in
Scott's book\cite{Scott00}. Typically  the currents observed
appear to be either emission-limited or space-charge-limited. In
this paper we restrict ourselves to the discussion of the
emission-limited type currents. What this paper shows is that a
metal-ferroelectric-metal system with space-charge due to oxygen
vacancies can be considered as a diffusion-limited current system.
The diffusion-limited current has two regimes: a low-voltage
regime akin to the low-voltage regime discussed in the theory of
semiconductor punch-through diodes\cite{Sze71}; and a high-voltage
regime equivalent to the diffusion-limited Schottky relationship
of Simmons\cite{Simmons65}.

\begin{figure}[h]
  \includegraphics[width=8.5cm]{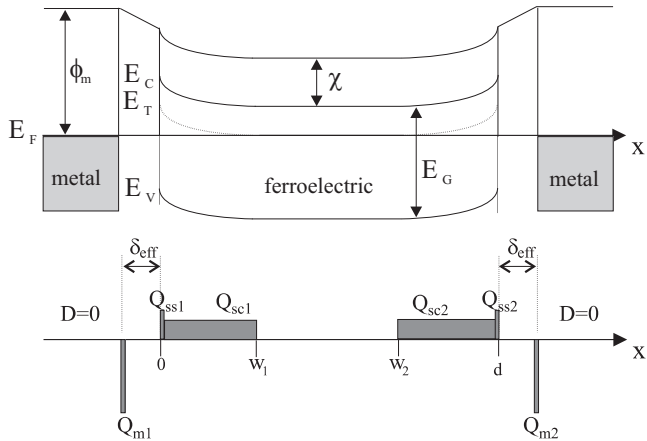}
  \caption{\textit{Band diagram and representation
of charges used to calculate field and potential
distributions}}\label{fig:chargediagram}
\end{figure}

We consider the field distribution in a capacitor (Fig. 1) which,
as well as having a large dielectric constant $\varepsilon_{f}$
and high uniform concentration of stationary oxygen vacancies
$N_{D}$, has a remnant polarisation, $P_{r}$. The thomas-fermi
screening length in the electrode and the penetration of the metal
states into the ferroelectric are accounted for by separating
$Q_{m}$, the charge on the metal from $Q_{ss}$, the charge due to
surface states (metal induced gap states) by an effective
thickness $\delta_{eff}$. This effective thickness reflects the
fact that there is a finite screening length in the metal (the
Thomas-Fermi screening length $\lambda_{tf}$) and that
metal-induced gap states penetrate a small distance into the
insulator ($\lambda_{MIGS}$).

\begin{equation}
\delta_{eff}=\frac{\lambda_{tf}}{\varepsilon_{0}}+\frac{\lambda_{MIGS}}{\varepsilon_{s
}\varepsilon_{0}}
\end{equation}

Metal induced gap states have previously have been shown to be
important in ferroelectric thin film capacitors by Robertson and
Chen\cite{Robertson99}. The metal screening length term is
typically the larger of the two, so $\delta_{eff}$ is usually
between about 0.5 and 1 $\frac{\AA}{\varepsilon_{0}}$

We now consider that the electric displacement in the film is
given by

\begin{equation}
D=\varepsilon_{f}\varepsilon_{0}E \pm P_{r}
\end{equation}

The sum of the charges in the system must equal zero since the
electric displacement is zero in both electrodes (far from the
interface). In the center of the film the field must equal the
field that would arise if there were no surface states or space
charge, which we take to be the linear drop $(\frac{\partial
V}{\partial x}=\frac{V}{d})$ that would occur if the sample were a
perfect insulator. This is quite different from the approach often
taken, which is to treat a metal-semiconductor-metal system as two
semi-infinite junctions, with the electric field taken as zero in
the regions of the film where there is no space charge, regardless
of the applied field.  Such an approach would be incompatible with
the space-charge-free limit of a linear voltage drop for our
insulating system. (This approach is however perfectly appropriate
for a single infinite-metal/infinite-semiconductor junction). In
addition, because our system is ferroelectric, when there is a
spontaneous polarisation a finite potential drop across the system
occurs because of the imperfect screening of the depolarisation
field, even in the absence of an applied field. Using these
arguments one finds that the field in the space-charge-free
interior of the film is

\begin{equation}
E_{0}=-\frac{1}{\varepsilon_{f}\varepsilon_{0}}\frac{V\pm
P_{r}2\delta_{eff}}{2\delta_{eff}+\frac{d}{\varepsilon_{f}\varepsilon_{0}}}
\end{equation}

Our expressions for the charge on the metal electrodes are then

\begin{equation}
Qm_{1}=-Q_{ss1}-qN_{D}w_{1}-\frac{V\mp
P_{r}\frac{d}{\varepsilon_{f}\varepsilon_{0}}}{2\delta_{eff}+\frac{d}{\varepsilon_{f}\varepsilon_{0}}}
\end{equation}

\begin{equation}
Qm_{2}=-Q_{ss2}-qN_{D}w_{2}+\frac{V\mp
P_{r}\frac{d}{\varepsilon_{f}\varepsilon_{0}}}{2\delta_{eff}+\frac{d}{\varepsilon_{f}\varepsilon_{0}}}
\end{equation}

The amount of charge due to surface states is independent of
voltage and is found by the condition that the fermi level at the
interface of the ferroelectric is
$S(\phi_{m}))+(1-S)(\chi+(E_{c}-\phi))$ below the vacuum level
where $S=\frac{1}{1+q^{2}\delta_{eff} D_{s}}$ and $\phi$ is the
charge neutrality level; this gives

\begin{equation}
Q_{ss1}=(1-S)(-qN_{D}w_{1}+\frac{1}{\delta_{eff}}((\phi_{m}-\chi)-(E_{g}-\phi)))
\end{equation}

This result shows that the charge due to surface states is
independent of voltage, and under the assumptions that the two electrodes are
the same and that the film is uniform, $Q_{ss1}=Q_{ss2}$.

Finally, the width of the depletion layers is found using the
condition that the potential at $x=w_{1}$ must be equal to the
height of the donor trap level $E_{t}$ (which in samples with low
intrinsic conductivity and high dopant levels will be the
semiconductor Fermi level). It is found that the depletion regions
are invariant with applied voltage,

\lefteqn{w_{1}+\varepsilon_{f}\varepsilon_{0}S\delta_{eff}=[((\varepsilon_{f}\varepsilon_{0}S\delta_{eff})^2+\frac{2\varepsilon_{f}\varepsilon_{0}}{qN_{D}}((\phi_{m}-\chi)}

\begin{equation}
-(E_{c}-E_{t})-(1-S)((\phi_{m}-\chi)-(E_{c}-\phi))]^{\frac{1}{2}}
\end{equation}

and that for a symmetrical system $w_{2}=w_{1}$. If a film is
fully depleted [i.e., if the calculated values of $w_{1}$ and $w_{2}$
are greater than $\frac{d}{2}$ (which means that the system fermi
level never reaches the trap level)], the following equations may
be used with $w_{1}=w_{2}=\frac{d}{2}$.

The finding that the depletion widths do not change with field is
at first quite surprising, the typical idea being that in
punch-through diodes, depletion widths will change with field and
that at a certain applied bias the reverse-biased depletion width
will ``punch-through'' to the forward-biased Schottky barrier. But
in fact what we show here is that it not the depletion width
itself that changes size, but rather the physical location within
the system of the field inversion point. It is when this field
inversion point coincides with the forward-biased Schottky barrier
that punch-through behaviour occurs.

The potential distributions for $0\leq x \leq w_{1}$,
$w_{1}<x<d-w_{2}$ and $d-w_{2}<x<d$ respectively are V(x)=

\begin{eqnarray}
\nonumber
Q_{m1}\delta_{eff}-E_{0}\frac{x}{\varepsilon_{f}\varepsilon_{0}}+\frac{qN_{D}}{\varepsilon_{f}\varepsilon_{0}}x(w_{1}-\frac{x}{2}),\\
\nonumber
Q_{m1}\delta_{eff}-E_{0}\frac{x}{\varepsilon_{f}\varepsilon_{0}}+\frac{qN_{D}}{2\varepsilon_{f}\varepsilon_{0}}w_{1}^{2},\\
\nonumber
Q_{m1}\delta_{eff}-E_{0}\frac{x}{\varepsilon_{f}\varepsilon_{0}}+\frac{qN_{D}}{2\varepsilon_{f}\varepsilon_{0}}w_{1}^{2}-\frac{qN_{D}}{2\varepsilon_{f}\varepsilon_{0}}(x-(d-w_{2}))^2\\
\end{eqnarray}

which give the field distributions in the same regions as

\begin{eqnarray}
\nonumber E(x)=E_{0}-\frac{qN_{D}}{\varepsilon_{f}\varepsilon_{0}}(w_{1}-x), \\
\nonumber E(x)=E_{0}, \\
\nonumber E(x)=E_{0}+\frac{qN_{D}}{\varepsilon_{f}\varepsilon_{0}}(x-(d-w_{2}) \\
\end{eqnarray}

An important result of this system of equations is that below a
certain critical voltage there is a point

\begin{equation}
x_{0}=-\frac{\varepsilon_{f}\varepsilon_{0}}{qN_{D}}E_{0}+d-w_{2}
\end{equation}

in the film at which the field is zero (the field inversion point
referred to above). We can calculate the current flowing in the
device by determining the point in the film at which the field is
zero. Here the current is entirely due to diffusion and is given
by

\begin{equation}
J=\mu kT \frac{\partial n}{\partial x}
\end{equation}

To evaluate $\frac{\partial n}{\partial x}$ we look at the number
of conduction electrons in the region $d-w_{2}<x<d$,

\begin{equation}
n(x)=N_{C}\exp(-\frac{q}{kT}((E_{c}-E_{t})+\frac{qN_{D}}{2\varepsilon_{f}\varepsilon_{0}}(x-(d-w_{2}))^2))
\end{equation}

Taking the derivative and using the solution for $x_{0}$ from
above we find the current to be

\begin{equation}
J=\mu N_{C}
E_{0}\exp(-\frac{q}{kT}((E_{c}-E_{t})+\frac{\varepsilon_{f}\varepsilon_{0}}{2qN_{D}}(E_{0})^2))
\end{equation}

An equivalent expression for the current, which is also applicable
for fully depleted films (in contrast to the earlier expression, which applies
to only partially depleted films) is

\begin{equation}
J=\mu N_{C} E_{0}\exp(-\frac{q}{kT}(\phi_{b}-\frac{
qN_{D}w^{2}}{2\varepsilon_{f}\varepsilon_{0}}+\frac{\varepsilon_{f}\varepsilon_{0}}{2qN_{D}}(E_{0})^2))
\end{equation}

For a fully depleted film one simply substitutes $w=\frac{d}{2}$
into the equation above. These relationships are applicable until
$x_{0}=d$ which is when
$\frac{V}{2\delta_{eff}\varepsilon_{f}\varepsilon_{0}+d}=qN_{D}w_{2}$
[or in other words when the field at the cathode is zero]. This is
equivalent to punch-through, but a punch-through of the field
inversion point, rather than the edge of the depletion width; and
it marks the transition to a new conduction regime. The current at
which the transition occurs is the same as Frank and Simmons
obtain\cite{Frank67}, i.e.,

\begin{equation}
J_{T}=-\mu N_{C}
\frac{qN_{D}w_{2}}{\varepsilon_{f}\varepsilon_{0}}
\exp(-\frac{\phi_{b}}{kT})
\end{equation}

The voltage at which the transition occurs,

\begin{equation}
V=qN_{D}w_{2}(2\delta_{eff}+\frac{d}{\varepsilon_{f}\varepsilon_{0}})
\end{equation}

is temperature- and thickness-dependent, the temperature
dependence is somewhat complicated because of the temperature
dependence of the dielectric constant (and hence also the
depletion width). When the applied voltage is increased beyond
this voltage, the image force pushes the potential maximum back
into the film, lowering the effective barrier. If we assume that
the new potential maximum is quite close to the interface, we can
use the field at the interface to calculate the Schottky barrier
lowering as

\begin{equation}
\Delta
\phi_{b}=\sqrt{\frac{q}{4\pi\varepsilon_{s}\varepsilon_{0}}(-E_{0}-\frac{qN_{D}w_{2}}{\varepsilon_{f}\varepsilon_{0}})}
\end{equation}

The dielectric constant $\varepsilon_{s}$ used here is the optical
dielectric constant of the material and is used because electrons
passing through the Schottky barrier do so sufficiently quickly
that they do not polarise the lattice\cite{Scott99}.

In this regime the current is

\begin{equation}
J=-\mu N_{C} E_{0}\exp(-\frac{q}{kT}(\phi_{b}-\Delta_{\phi_{b}}))
\end{equation}

We note that this is the form of the Richardson-Schottky equation
applicable in materials with very short mean free paths and has
been previously derived by Simmons \cite{Simmons65}. This form of
the Richardson-Schottky equation has been already been noted as
the appropriate one for ferroelectric thin films and has been used
successfully to describe leakage current in BST thin films by
Zafar\cite{Zafar98b}

In a previous study on Au-BST-SrRuO$_{3}$
capacitors\cite{Dawber02} we incorrectly assumed that the high-
voltage regime corresponded to tunnelling through the space-charge
region in the electrode because of the correspondence between the
threshold for the high-voltage regime and the applied voltage at
which the potential drop across the electrode equalled the barrier
height; i.e., it was noted that

\begin{equation}
V_{T}=\frac{\phi_{b}}{2\delta_{eff}}(2\delta_{eff}+\frac{d}{\varepsilon_{f}\varepsilon_{0}})
\end{equation}

We note however that the present model predicts the same thickness
dependence of the threshold voltage:
$V\propto(2\delta_{eff}+\frac{d}{\varepsilon_{f}\varepsilon_{0}})$.
Low temperature measurements on the 70 nm film from this study
show that the high-voltage regime is Schottky-like both at room
temperature and at 70 K (Fig. 2).

\begin{figure}[h]
  \includegraphics[width=8.5cm]{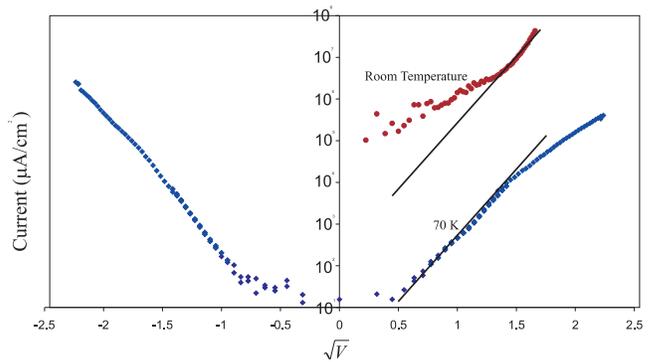}
  \caption{\textit{Leakage current data from 70nm thick Au-BST-SrRuO$_{3}$ film
at room temperature and at T=70 K}}\label{fig:fig3}
\end{figure}

The low-field expression derived above (Eqs. 15 and 16) can
account for two kinds of low-field behaviour observed in
ferroelectric thin films: either a sharp increase followed by a
heretofore unexplained flat plateau (which is the behavior of the
equation when the donor concentration is very high) or a negative
differential resistivity (observed for lower donor
concentrations).

\begin{figure}[h]
  \includegraphics[width=8.5cm]{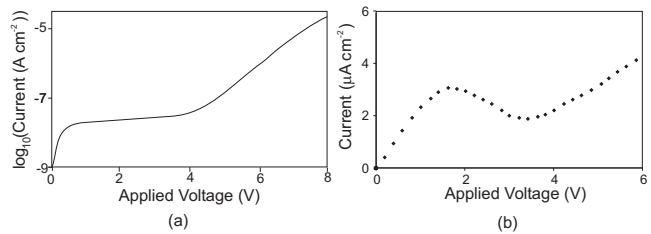}
  \caption{\textit{(a) Scott et al.\cite{Scott94}: Leakage current of BST
thin film showing flat plateau followed by Schottky regime, (b)
Watanabe et al.\cite{Watanabe98}: Leakage current of SBT film
showing negative differential resistivity followed by Schottky
regime}}\label{fig:lcfig}
\end{figure}

An example of the first kind of behavior (Fig. 3(a)) is found in
the data of Scott et al\cite{Scott94} on BST thin films; whereas a
good examples of the second (Fig. 3(b)) can be found in the SBT
samples of Watanabe et al\cite{Watanabe98} or in PZT in the data
of Scott et al.\cite{Scott94b} or Chen et al\cite{Chen97}.
Previous attempts to explain these effects have invoked the
filling and emptying of trap states.

On taking the derivative with respect to voltage of our low-field
expression, we find that

\lefteqn{\frac{\partial J}{\partial V}=\frac{\mu
N_{c}}{2\delta_{eff}
\varepsilon_{f}\varepsilon_{0}+d}(1-(E_{0})^2\frac{\varepsilon_{f}\varepsilon_{0}}{kT
N_{D}})\times}
\begin{equation}
\exp(-\frac{q}{kT}(\phi_{b}-\frac{
qN_{D}w^{2}}{2\varepsilon_{f}\varepsilon_{0}}+\frac{\varepsilon_{f}\varepsilon_{0}}{2qN_{D}}(E_{0})^2))
\end{equation}

The exponential term decreases with voltage towards the value
$\exp(-\frac{\phi_{b}}{kT})$ but is always positive. If the other
voltage-dependent term remains positive, the low-voltage leakage
current characteristic will be a rapid increase which plateaus off
into an ohmic regime until such voltage that the barrier begins to be
lowered by the Schottky effect. However, if the voltage is greater
than

\begin{equation}
V>(2\delta_{eff}\varepsilon_{f}\varepsilon_{0}+d)\sqrt{kT\varepsilon_{f}^{-1}\varepsilon_{0}^{-1}N_{D}}
\end{equation}

the differential resistivity will become negative. We should not
forget, however, the condition that this conduction mechanism only
operates when

\begin{equation}
V<2qN_{D}w_{2}(2\delta_{eff}+\frac{d}{\varepsilon_{f}\varepsilon_{0}})
\end{equation}

and so negative differential resistivity is observed only if

\begin{equation}
\sqrt{\frac{\varepsilon_{f}\varepsilon_{0}kT}{4q^{2}N_{D}w_{2}^2}}<1
\end{equation}

Within the Schottky injection regime one of the most interesting
things to note is the role that the dielectric constant plays in
determining the temperature dependence of the current. We can show
that when the sample is below a ferroelectric phase transition and
the dielectric constant is increasing rapidly with temperature a
regime of Positive Temperature Coefficient of Resistivity (PTCR)
may be observed. For this purpose we simplify the Schottky regime
current expression by taking

\begin{equation}
\Delta
\phi_{b}=\sqrt{\frac{q}{4\pi\varepsilon_{s}\varepsilon_{0}}(-E_{0})}
\end{equation}

and taking the derivative with respect to temperature we find that

\begin{equation}
\frac{\partial J}{\partial
T}=J(-\frac{q}{kT}[\frac{1}{T}(\phi_{b}-\Delta\phi_{b})-\Delta\phi_{b}(\frac{\partial
\varepsilon_{f}}{\partial
T}(\frac{\delta_{eff}}{2\delta_{eff}\varepsilon_{f}\varepsilon_{0}+d}))])
\end{equation}

The term $\frac{1}{T}(\phi_{b}-\Delta\phi_{b})$ is always
positive. For a PTCR effect to be observed above a certain applied
voltage $\frac{\partial \varepsilon_{f}\varepsilon_{0}}{\partial
T}(\frac{\delta_{eff}}{2\delta_{eff}\varepsilon_{f}\varepsilon_{0}+d})$
must be large and positive, and hence the effect should be seen
only below T$_{C}$. Whether or not PTCR behaviour is observed
depends on the barrier height and the screening length in the
metal, with PTCR behaviour more likely to occur for capacitors
with lower barrier heights and longer screening lengths.The PTCR
effect predicted here does not involve grain boundaries, as
distinct from the PTCR effect in BaTiO$_{3}$ ceramics (for a
review see Huybrechts et al. \cite{Huybrechts95}) which as shown
by Sinclair and West \cite{Sinclair89} has both bulk and grain
boundary contributions. Hwang\cite{HwangPTCR} has observed PTCR
effects in films with Ir electrodes, whereas similar films with Pt
electrodes did not display PTCR behaviour. This is consistent with
our model because Ir has a significantly lower workfunction than
Pt (4.23 eV compared to 5.3 eV \cite{Chalamala99}).

Thus with a single model we have explained many unusual leakage
current features in ferroelectric capacitors

\end{document}